\documentclass[aps,prl,reprint,superscriptaddress]{revtex4-2}
\usepackage{siunitx}
\usepackage{graphicx}
\usepackage[nopatch=footnote]{microtype}
\usepackage[hidelinks]{hyperref}
\usepackage{amsmath}
\newcommand{\fref}[1]{FIG.~\ref{#1}}
\newcommand{\aref}[1]{Appendix~\ref{#1}}
\newcommand{\Var}{\operatorname{Var}}

\begin{document}
\preprint{arXiv:2512.06118}
\title{Nonlinear phenomena in X-ray fluorescence from single nanoparticles under extreme conditions}
\author{Sebastian Cardoch}
\thanks{These authors contributed equally to this work.}
\affiliation{Department of Physics and Astronomy, Uppsala University, Box 516, Uppsala SE-75120, Sweden}
\author{Tamme Wollweber}
\thanks{These authors contributed equally to this work.}
\affiliation{Max Planck Institute for the Structure and Dynamics of Matter, 22761 Hamburg, Germany}
\affiliation{The Hamburg Centre for Ultrafast Imaging, 22761 Hamburg, Germany}
\author{Sarodi Jonak Dutta}
\affiliation{The Hamburg Centre for Ultrafast Imaging, 22761 Hamburg, Germany}
\affiliation{Department of Physics, Universität Hamburg, Luruper Chaussee 149, 22761 Hamburg, Germany}
\author{Zhou Shen}
\affiliation{Max Planck Institute for the Structure and Dynamics of Matter, 22761 Hamburg, Germany}
\author{Johan Bielecki}
\affiliation{European XFEL, Holzkoppel 4, 22869 Schenefeld, Germany}
\author{Fabian Trost}
\affiliation{European XFEL, Holzkoppel 4, 22869 Schenefeld, Germany}
\author{Armando D. Estillore}
\affiliation{Center for Free-Electron Laser Science CFEL, Deutsches Elektronen-Synchrotron DESY, Notkestr. 85, 22607 Hamburg, Germany}
\author{Lukas V. Haas}
\affiliation{The Hamburg Centre for Ultrafast Imaging, 22761 Hamburg, Germany}
\affiliation{Department of Physics, Universität Hamburg, Luruper Chaussee 149, 22761 Hamburg, Germany}
\affiliation{Center for Free-Electron Laser Science CFEL, Deutsches Elektronen-Synchrotron DESY, Notkestr. 85, 22607 Hamburg, Germany}
\author{Sebastian Karl}
\affiliation{Department of Physics, Friedrich-Alexander-Universität Erlangen-Nürnberg, Staudtstrasse 1, 91058 Erlangen, Germany}
\author{Faisal H.M. Koua}
\affiliation{European XFEL, Holzkoppel 4, 22869 Schenefeld, Germany}
\author{Abhishek Mall}
\affiliation{Max Planck Institute for the Structure and Dynamics of Matter, 22761 Hamburg, Germany}
\author{Parichita Mazumder}
\affiliation{Max Planck Institute for the Structure and Dynamics of Matter, 22761 Hamburg, Germany}
\author{Diogo Melo}
\affiliation{European XFEL, Holzkoppel 4, 22869 Schenefeld, Germany}
\author{Mauro Prasciolu}
\affiliation{Center for Free-Electron Laser Science CFEL, Deutsches Elektronen-Synchrotron DESY, Notkestr. 85, 22607 Hamburg, Germany}
\author{Omkar V. Rambadey}
\affiliation{Max Planck Institute for the Structure and Dynamics of Matter, 22761 Hamburg, Germany}
\author{Amit Kumar Samanta}
\affiliation{The Hamburg Centre for Ultrafast Imaging, 22761 Hamburg, Germany}
\affiliation{Center for Free-Electron Laser Science CFEL, Deutsches Elektronen-Synchrotron DESY, Notkestr. 85, 22607 Hamburg, Germany}
\author{Abhisakh Sarma}
\affiliation{European XFEL, Holzkoppel 4, 22869 Schenefeld, Germany}
\author{Tokushi Sato}
\affiliation{European XFEL, Holzkoppel 4, 22869 Schenefeld, Germany}
\author{Egor Sobolev}
\affiliation{European XFEL, Holzkoppel 4, 22869 Schenefeld, Germany}
\author{Sasa Bajt}
\affiliation{The Hamburg Centre for Ultrafast Imaging, 22761 Hamburg, Germany}
\affiliation{Center for Free-Electron Laser Science CFEL, Deutsches Elektronen-Synchrotron DESY, Notkestr. 85, 22607 Hamburg, Germany}
\author{Richard Bean}
\affiliation{European XFEL, Holzkoppel 4, 22869 Schenefeld, Germany}
\author{Carl Caleman}
\affiliation{Department of Physics and Astronomy, Uppsala University, Box 516, Uppsala SE-75120, Sweden}
\affiliation{Center for Free-Electron Laser Science CFEL, Deutsches Elektronen-Synchrotron DESY, Notkestr. 85, 22607 Hamburg, Germany}
\author{Jochen K{\"u}pper}
\affiliation{The Hamburg Centre for Ultrafast Imaging, 22761 Hamburg, Germany}
\affiliation{Department of Physics, Universität Hamburg, Luruper Chaussee 149, 22761 Hamburg, Germany}
\affiliation{Center for Free-Electron Laser Science CFEL, Deutsches Elektronen-Synchrotron DESY, Notkestr. 85, 22607 Hamburg, Germany}
\affiliation{Department of Chemistry, Universität Hamburg, 20146 Hamburg, Germany}
\author{Ralf R{\"o}hlsberger}
\affiliation{The Hamburg Centre for Ultrafast Imaging, 22761 Hamburg, Germany}
\affiliation{Deutsches Elektronen-Synchrotron DESY, Notkestr. 85, 22607 Hamburg, Germany}
\affiliation{Helmholtz-Institut Jena, Fröbelstieg 3, 07743 Jena, Germany}
\affiliation{GSI Helmholtzzentrum für Schwerionenforschung, Planckstrasse 1, 62491 Jena, Germany}
\affiliation{Institut für Optik und Quantenelektronik, Friedrich-Schiller-Universität Jena, Max-Wien-Platz 1, 07743 Jena, Germany}
\author{Joachim von Zanthier}
\affiliation{Department of Physics, Friedrich-Alexander-Universität Erlangen-Nürnberg, Staudtstrasse 1, 91058 Erlangen, Germany}
\affiliation{Erlangen Graduate School in Advanced Optical Technologies (SAOT), Friedrich-Alexander Universität Erlangen-Nürnberg, Paul-Gordan-Str. 6, 91052, Erlangen, Germany}
\author{Florian Schulz}
\affiliation{The Hamburg Centre for Ultrafast Imaging, 22761 Hamburg, Germany}
\affiliation{Department of Physics, Universität Hamburg, Luruper Chaussee 149, 22761 Hamburg, Germany}
\author{Henry N. Chapman}
\affiliation{The Hamburg Centre for Ultrafast Imaging, 22761 Hamburg, Germany}
\affiliation{Department of Physics, Universität Hamburg, Luruper Chaussee 149, 22761 Hamburg, Germany}
\affiliation{Center for Free-Electron Laser Science CFEL, Deutsches Elektronen-Synchrotron DESY, Notkestr. 85, 22607 Hamburg, Germany}
\author{Kartik Ayyer}
\email{kartik.ayyer@mpsd.mpg.de}
\affiliation{Max Planck Institute for the Structure and Dynamics of Matter, 22761 Hamburg, Germany}
\affiliation{The Hamburg Centre for Ultrafast Imaging, 22761 Hamburg, Germany}
\author{Nicusor Timneanu}
\email{nicusor.timneanu@physics.uu.se}
\affiliation{Department of Physics and Astronomy, Uppsala University, Box 516, Uppsala SE-75120, Sweden}
\date{\today}
\begin{abstract}
  Materials exposed to intense femtosecond X-ray pulses with energies above their K-shell absorption edge can enter an extremely ionized state, which could give rise to nonlinear phenomena, such as saturable absorption and reverse saturable absorption. In this work, we investigate these effects on single copper nanoparticles irradiated by an X-ray free-electron laser pulse. We study the properties of the K$\alpha$ fluorescence for two different pulse durations and three X-ray incident energies below and above the K-shell absorption edge, and correlate these with incident fluence estimates based on coherent diffraction. We observe that the incident fluence of the pulse and not its duration, is the main factor that modulates the nonlinear response, which leads to an effective shortening of the fluorescence emission. Our findings have implications for fluorescence-based methods for imaging single particles using transiently coherent fluorescence, or diffractive imaging through transient resonances.
\end{abstract}
\maketitle
X-ray free-electron laser (XFEL) sources featuring ultra-short pulse durations and unprecedented peak brilliance have prompted new opportunities for investigating light-matter interactions under extreme conditions~\cite{rudek2018relativistic,ho2020role}. A striking example is the method of incoherent diffractive imaging (IDI) that uses intensity correlations of X-ray fluorescence for imaging. In the preceding Letter~\cite{wollweber2025} to this manuscript, we demonstrate single-particle IDI and show that IDI visibility collapses at high fluence. Here, using the same XFEL experiment, we characterize the underlying nonlinear X-ray fluorescence and self-gating of single nanoparticles that underpin these observations.

At very high fluence and on timescales on the order of a few femtoseconds, multiple photons can interact with a single atom during an XFEL pulse~\cite{young2010femtosecond}. For X-ray pulses with energies above the K-shell absorption edge, the primary light-matter interaction involves \(1s\) photoionization. The additional energy imparted to the atomic system is then released via competing radiative and nonradiative mechanisms such as X-ray fluorescence, Auger-Meitner decay, and shake events. These relaxation channels take place a few femtoseconds or less after photoionization~\cite{young2010femtosecond,rohringer2007x}. Alongside primary ionization, electron impact from free electrons generates further vacancies in the material via the depopulation of outer valence states of the whole atomic ensemble~\cite{caleman2011feasibility}.

Thus, intense femtosecond X-ray pulses with photon energies above absorption edges can drive matter into transient, highly ionized states in which the X-ray response becomes strongly nonlinear. In bulk and extended samples, this regime has been shown to give rise to \emph{saturable absorption}~\cite{nagler_turning_2009,yoneda2014saturable,rackstraw2015saturable,hoffmann_saturable_2022} and \emph{reverse saturable absorption}~\cite{cho2017observation,mercadier2024transient}, transient resonances that enhance ultrafast diffraction~\cite{kuschel_nonlinear_2025} and enable self-gating of the X-ray emission~\cite{inoue2021shortening}. Yet it remains largely unexplored how these effects manifest at the level of single nanoparticles, where the finite size and short gain length would support the naive expectation that collective phenomena are suppressed.

Studies based on X-ray pulse transmission or absorption spectra, and supported by simulations, reveal that absorption changes occur when photons tuned to photoionize a specific shell severely deplete the shell's occupation and outrun electronic relaxation mechanisms. In addition, electron impact valence ionization leads to an increase in the binding energy of remaining bound electrons, shifting the absorption edge to higher energies~\cite{son2011multiwavelength}. For example, when exposed to photon energies just above the K-shell, the material becomes transparent as the absorption edge is suppressed and surpasses the incident pulse energy (saturable absorption). The reverse behavior can be observed when exposing the material to a photon energy slightly below the K-shell absorption edge. Although these photons are insufficiently energetic to ionize \(1s\) electrons, they can photoionize from L, M, or higher shells. As a result, the energy difference between the K- and the M-shells increases until this transition becomes resonant with the incident photon energy~\cite{cardoch2023decreasing}. This resonance causes a sudden increase in the absorption cross section, rendering the material opaque (reverse saturable absorption).

In this letter, we report the emergence of these two nonlinear effects in an XFEL experiment on individual copper nanoparticles based on their K\(\alpha\) fluorescence. We scanned across the vicinity of the copper's K-edge (\SI{8978.9}{\electronvolt}~\cite{henke1993x}) with three different photon energies \SI{8910}{\electronvolt} (below the edge), \SI{9015}{\electronvolt} (just above the edge), and \SI{9170}{\electronvolt} (well above the edge). The measurements were conducted for two different XFEL operation modes, providing long and short pulses with nominal pulse durations and energies of \SI{25}{\femto\second} with \SI{2}{\milli\joule}, and \SI{2.5}{\femto\second} with \SI{200}{\micro\joule}, respectively.

\begin{figure}
  \includegraphics[width=\linewidth]{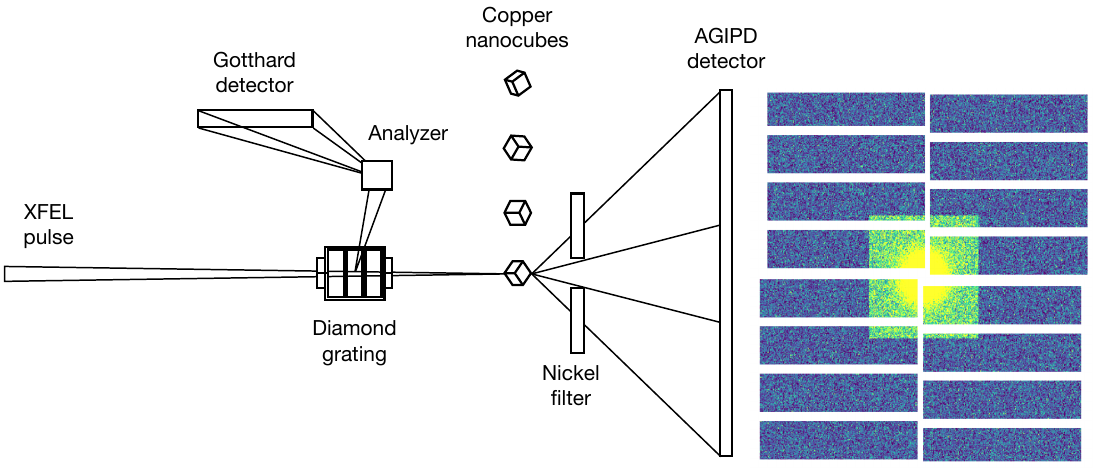}
  \caption{\label{fig:schematic}Schematic illustration of the experimental setup. The spectrum of each incident XFEL pulse is recorded by a hard X-ray single-shot spectrometer~\cite{hirex} using the Gotthard detector~\cite{amozzanica_gotthard_2012}. The copper nanocubes are aerosolized and delivered to the interaction region, where they are intercepted by the X-rays. Both scattering and fluorescence are recorded in the forward region by the AGIPD detector~\cite{becker_single_2012}. The Nickel filter allows coherent diffraction at low angles because of an opening around the direct beam, which can be used for size selection, orientation determination, and classification~\cite{wollweber2025}. At high angles, elastic photons are filtered out, and K\(\alpha\) fluorescence is predominant on the detector (right panel).}
\end{figure}

A diagram of the experimental setup is shown in \fref{fig:schematic}. X-ray pulses intercept individual aerosolized copper nanoparticles approximately \SI{88}{\nano\meter} in diameter~\cite{schulz}. Along the forward direction using the Adaptive Gain Integrating Pixel Detector (AGIPD)~\cite{becker_single_2012}, we concurrently record coherent diffraction in the central region of the detector and copper K\(\alpha\) fluorescence in the outer regions of the detector. This arrangement is possible by placing a \SI{20}{\micro\meter}-thick nickel foil (K-edge at \SI{8332}{\electronvolt}~\cite{henke1993x}) between the interaction region and detector. A \SI{10}{\milli\meter} by \SI{10}{\milli\meter} square center cutout on the foil allowed all photon energies to be recorded, including coherent diffraction which will be used to classify the size and orientation of the nanoparticles. The remaining foil creates a shadow over the detector by absorbing elastically scattered photons and copper K\(\beta\) fluorescence, while allowing K\(\alpha\) fluorescence to pass through. The photon energy distribution of each pulse was recorded using a hard X-ray single-shot spectrometer~\cite{hirex} using a Gotthard detector~\cite{amozzanica_gotthard_2012}. This single-particle approach allows us to uniquely study light-matter interactions on a scale where collective phenomena such as amplified spontaneous emission (ASE)~\cite{rohringer2012atomic,yoneda2015atomic,benediktovitch2020amplified} are not naively expected due to the short gain length.

\begin{figure}[ht]
  \centering
  \includegraphics[width=\linewidth]{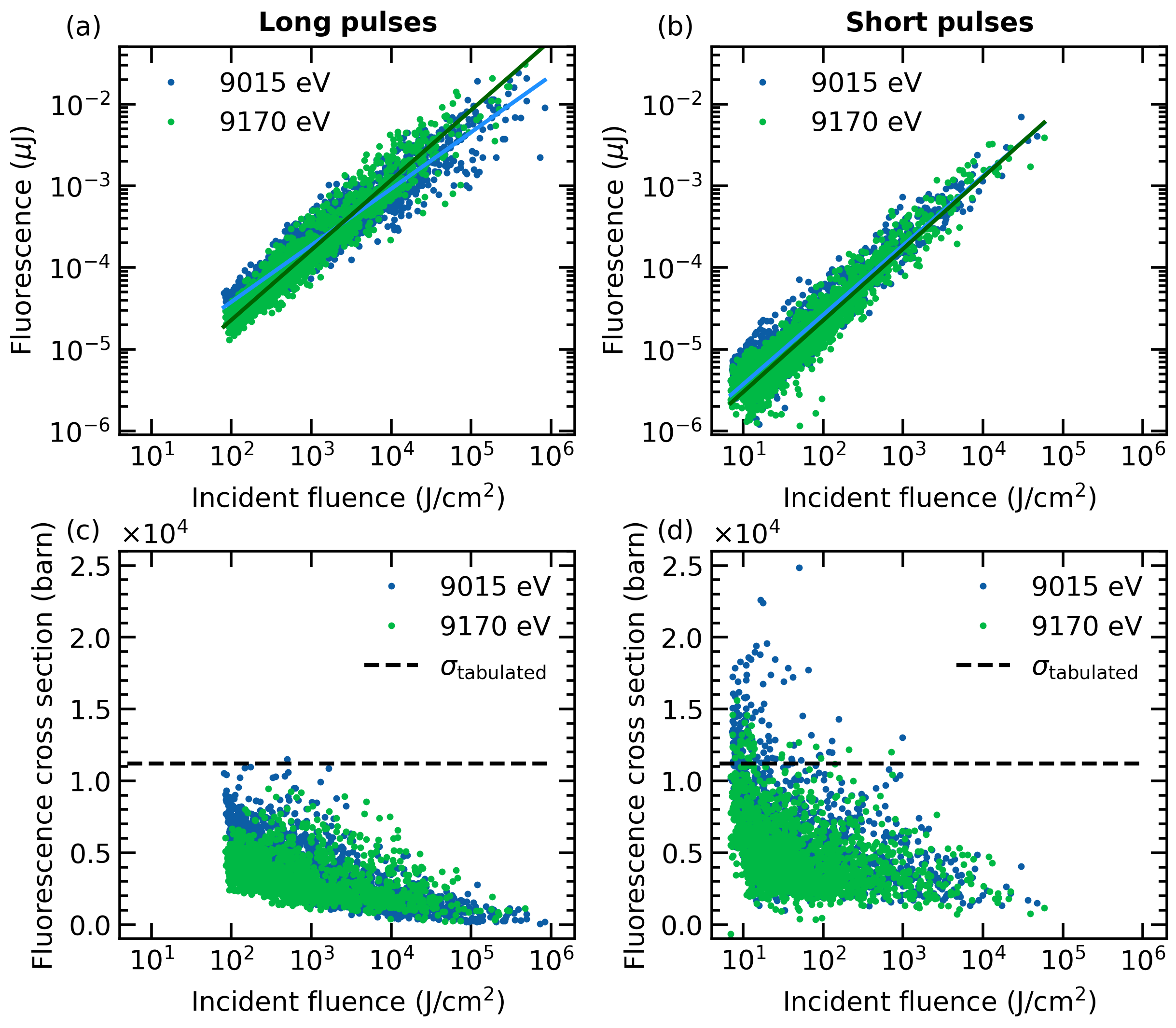}
  \caption{\label{fig:exp_above_edge}Signature of saturable absorption on single copper nanoparticles for \SI{25}{\femto\second} (long pulses, left column) and \SI{2.5}{\femto\second} (short pulses, right column) pulses as a function of incident fluence for \SI{9015}{\electronvolt} and \SI{9170}{\electronvolt} photon energies. (a,b)~Aggregate of single-shot fluorescence over the incident fluence values, determined by the total intensity on the region shadowed by the nickel filter and the region exposed by the foil cutout, respectively. We trace a linear fit (solid line) for each photon energy, which reveals a crossover between the two energies. (c,d)~Fluorescence cross sections based on the fluorescence intensity, from single particle measurements. The horizontal lines show the tabulated cross section.}
\end{figure}

We first examine the total fluorescence from individual patterns at excitation energies above the edge. Single hits are determined after filtering for misses and performing 2D classification using X-ray single particle imaging methods. The incident X-ray fluence is determined by fitting the coherently scattered intensity over the detector region exposed by the foil cutout and transformed to absolute units based on tabulated cross sections (see~\cite{wollweber2025} for more details about the coherent diffraction analysis). The position of the individual particles relative to the X-ray focus varies from pulse to pulse. This gives us access to a broad range of incident fluence values. We determine the total copper K\(\alpha\) emission by integrating the intensity over the region shadowed by the nickel, scaling by the filter's transmission, and detector solid angle. The results for incident photon energies $\SI{9015}{\electronvolt}$ and $\SI{9170}{\electronvolt}$ as well as for long and short pulse durations, are shown in \fref{fig:exp_above_edge}(a,b). Under a linear response of matter, we expect a higher fluorescence emission from X-rays with a photon energy nearest to the K-edge~\cite{mcmorrow_elements_2011}. Based on a linear best fit to the data (solid lines), we find this holds for fluence values \SI{<1E3}{\joule\per\centi\meter\squared}. As the incident fluence increases, fluorescence from \SI{9170}{\electronvolt} becomes stronger, suggesting changes in the absorption.

The signature of saturable absorption is most apparent when comparing fluorescence cross sections as shown in \fref{fig:exp_above_edge}(c,d). Details for determining these values based on fluorescence are provided in \aref{sec:sigma}. For both pulse durations, the absorption cross section at low fluence values approaches the cold copper fluorescence absorption. With increasing fluence, the cross-section decreases substantially. While the absorption behavior for the two excitation energies is qualitatively similar, the biggest differences occur for the highest fluence values and the longer pulse durations.

\begin{figure}
  \centering
  \includegraphics[width=\linewidth]{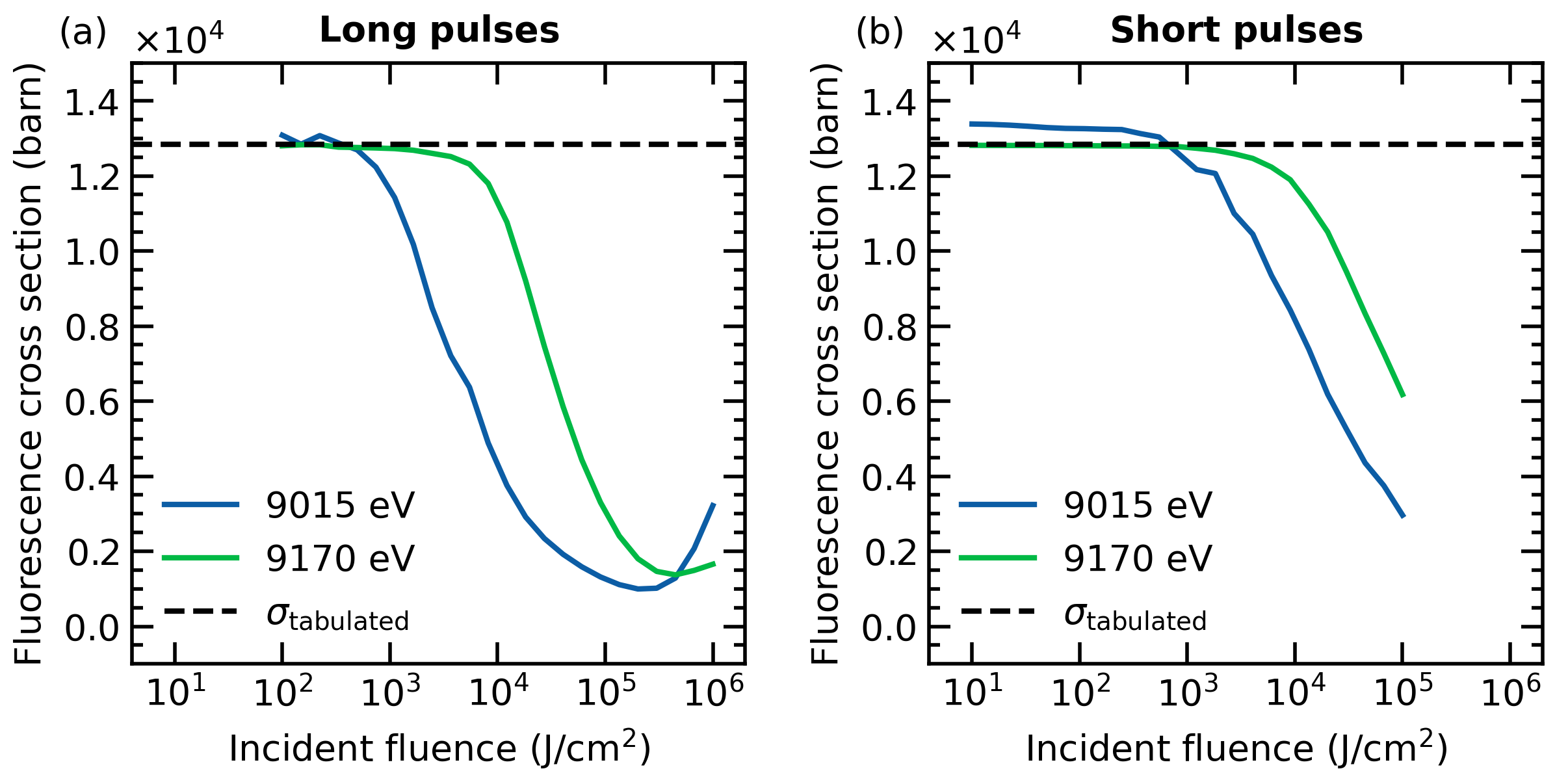}
  \caption{\label{fig:th_cross_section}Simulated copper fluorescence cross section as a function of incident X-ray fluence, for the (a)~\qty{25}{\femto\second}, and (b)~\qty{2.5}{\femto\second} X-ray pulses. The cross sections are based on pulse-weighted changes to the linear absorption coefficient at \SI{9015}{\electronvolt} and \SI{9170}{\electronvolt}.}
\end{figure}

We theoretically examine this phenomenon with a collisional radiative code~\cite{scott_glf_1994,scott_cretinradiative_2001}, together with relativistic configuration-averaged atomic data calculated using the Flexible Atomic Code~\cite{gu_flexible_2008}. The simulations include the interaction with the X-rays and electron impact processes. The code has been employed in several XFEL studies of solids, liquids, and protein crystals~\cite{barty_selfterminating_2012,beyerlein_ultrafast_2018,cardoch2023decreasing}. In the simulations, we assume the incident X-rays are modeled by a Gaussian function in the time domain with \SI{25}{\femto\second} and \SI{2.5}{\femto\second} full width at half maximum for long and short pulses, respectively. The copper nanoparticles are modeled as a one-dimensional \SI{88}{\nano\meter}-thick system with a density of \SI{8.96}{\gram\per\centi\meter\cubed} and initial temperature of \SI{0.025}{\electronvolt}. The bandwidth of the X-rays is assumed to be \SI{0.3}{\percent} flat-top shaped. Additional simulation details are given in \aref{sec:modelling}. In \fref{fig:th_cross_section}, we plot the pulse-weighted fluorescence cross section at different incident X-ray fluence values. Similar to experimental results, the fluorescence cross section decreases substantially at the highest fluence values, with cross sections at \SI{9015}{\electronvolt} excitation energy falling at earlier fluences compared to \SI{9170}{\electronvolt} excitation energy. The simulations estimate the fluence for the onset of saturation \SI{\sim1E3}{\joule\per\centi\meter\squared}, which the experiment suggests takes place \SI{\sim1E2}{\joule\per\centi\meter\squared}. This disagreement may be partly due to the approximate method used for simulating the pulse duration and bandwidth. Another source of uncertainty arises from nonlinear resonantly enhanced coherent scattering that can affect the experimentally determined incident fluences~\cite{kuschel_nonlinear_2025}. Although this effect cannot explain the discrepancy in the onset of transparency between experiment and theory, it could modify the cutoff incident fluence.

\begin{figure}
  \centering
  \includegraphics[width=\linewidth]{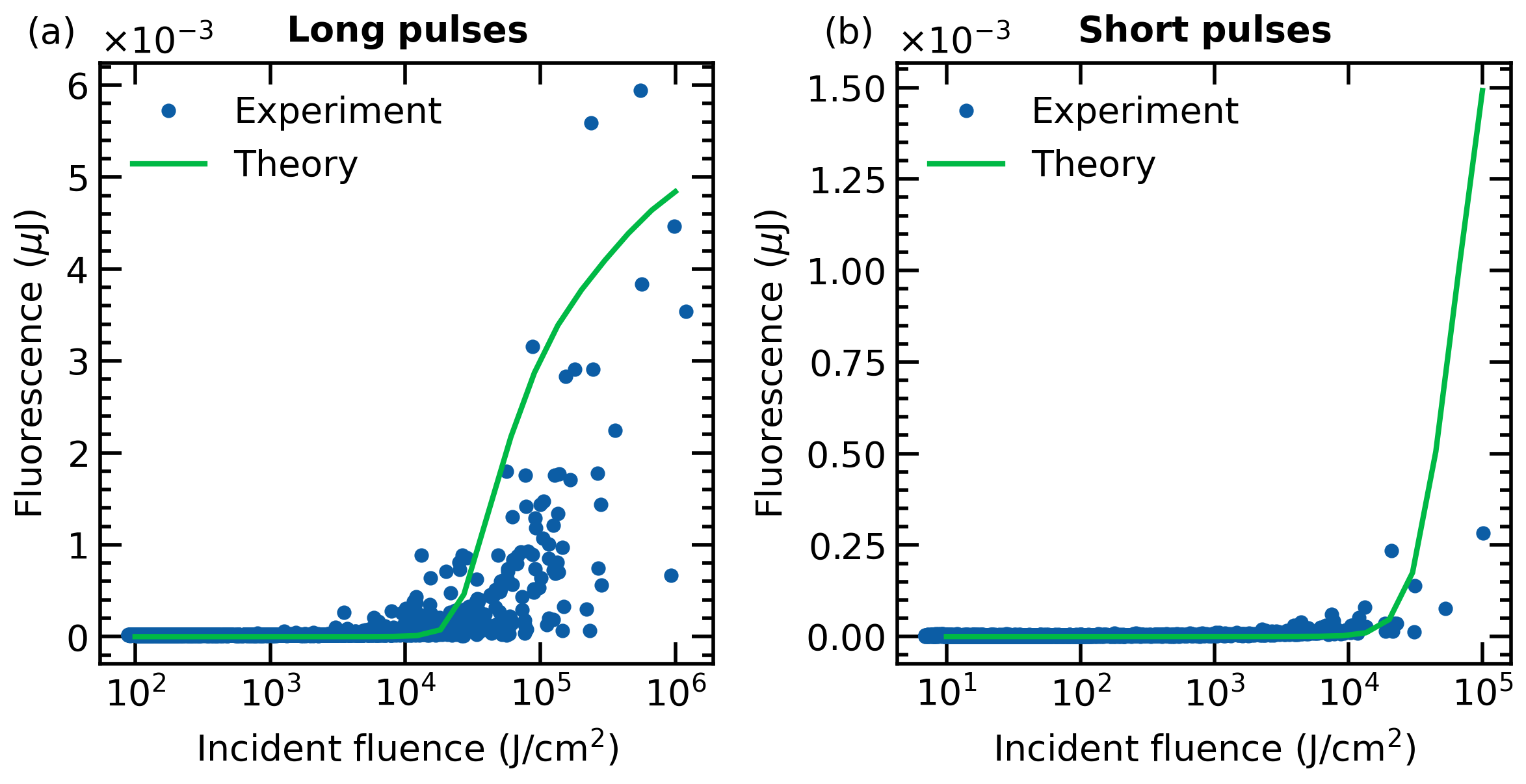}
  \caption{\label{fig:below_edge}Signature of reverse saturable absorption on single copper nanoparticles. Experimental and theoretical copper K\(\alpha\) fluorescence from (a)~\SI{25}{\femto\second} and (b)~\SI{2.5}{\femto\second} duration incident X-ray pulses with \SI{8910}{\electronvolt} photon energies, below the copper K-shell absorption edge.}
\end{figure}

We turn our attention next to fluorescence emission from nanoparticles exposed to \SI{8910}{\electronvolt} pulses below the copper K-shell absorption edge. Based on tabulated values, we expect the absorption cross section to be minimal and fluorescence to be negligible. However, when the incident fluence is high enough, the excitation of lower-bound states induces M-shell vacancies and the K-to-M-shell transition (\SI{8905}{\electronvolt}) becomes resonant with the excitation pulse energy, thereby leading to a sudden increase of the absorption cross section and resulting fluorescent emission~\cite{cardoch2023decreasing}. This is demonstrated in \fref{fig:below_edge}, with points corresponding to experimental observations and the line corresponding to simulations. Theoretical results include \SI{43}{\percent} transmission from the nickel foil as well as a \SI{6}{\percent} solid angle and \SI{95}{\percent} quantum efficiency~\cite{becker_single_2012} of the detector.

Since the photon energy is below copper's original K-edge, this fluorescence can only be attributed to the K-M-shell resonance. The effect is even more pronounced for longer pulses, where the incident fluence is approximately ten times higher than for the shorter pulses. In both cases, the resonance appears at the same incident fluence \SIrange{E3}{E4}{\joule/\centi\meter\squared}, indicating that on femtosecond time scales, the resonance is governed by the energy deposited in the particle rather than the pulse duration. Note that the resonant emission happens in the same fluence region where the cross-over happens for saturable absorption, suggesting a similar underlying response of lower-bound electronic states.

\begin{figure}
  \centering
  \includegraphics[width=\linewidth]{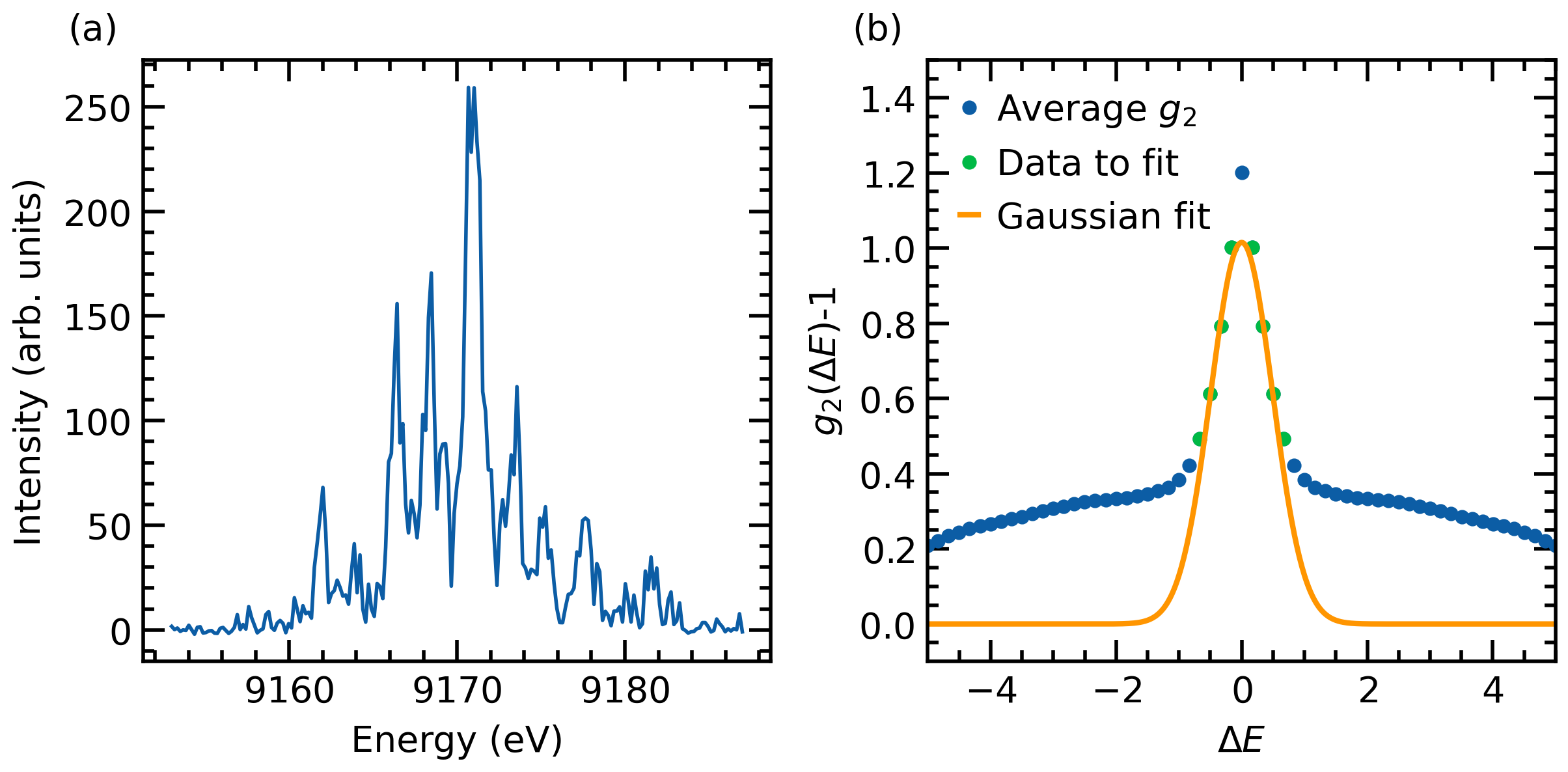}
  \caption{Estimation of the short pulse duration. (a)~Exemplary hard X-ray single-shot spectrum. (b) Second order correlation function~$g^{(2)}(\Delta E)$ for all spectra with a photon energy of \SI{9170}{\electronvolt}. The full width at half maximum of the central feature is inversely proportional to the pulse duration and of the broad feature (fit not shown) to the averaged single SASE spike duration.}
  \label{fig:hirex}
\end{figure}

Our experiment also gives us a unique opportunity to directly measure the short pulse duration. An example of the energy distribution of each pulse recorded with a hard X-ray single-shot spectrometer is shown in \fref{fig:hirex}(a). The spectra of self-amplified spontaneous emission (SASE) pulses exhibit shot-to-shot fluctuations, affecting both the center frequency, the local spike width, and the relative intensity profile within an average Gaussian envelope. To estimate the average spike width, which is inversely proportional to the temporal duration of the XFEL pulse~\cite{yan2024terawatt}, we calculate the second-order correlation function \(g^{(2)}(\Delta E) = \langle I(E) \cdot I(E+\Delta E) \rangle/\langle I(E) \rangle^2\) on the spectra. The resulting $g^{(2)}(\Delta E)$, shown in \fref{fig:hirex}(b) based on all spectra at \SI{9170}{\electronvolt}, contains two features corresponding to the average spike and overall envelope widths. A Gaussian fit to the central feature yields an average spike width of $\Delta E=\SI{0.65}{\electronvolt}$. Assuming a transformed-limited pulse duration and based on Equation~1 of \cite{huang_generating_2017}, {we obtain an estimate of \SI{2.8}{\femto\second}}. Carrying out a similar procedure for the broader feature returns an average single spike width of \SI{9}{\electronvolt}, corresponding to a pulse duration of \SI{\sim 200}{\atto\second}, which is in good agreement with results presented in~\cite{yan2024terawatt}. For long pulses, peaks in the spectra are closely packed and too narrow for the \(g_{2}(\Delta E)\) to yield a reasonable estimate.

Frustrated absorption and the increased transparency should lead to a reduction of the effective pulse duration of the excitation pulse~\cite{inoue2021shortening,cardoch2023decreasing}. The nature of the emitted fluorescence can be used to independently estimate the extent of this so-called self-gating effect, using the internal clock of the fluorescence coherence time of \(\tau_\mathrm{c}=\SI{0.49}{\femto\second}\)~\cite{holzer1997k}. 
The photon statistics of partially coherent light such as $K\alpha$ fluorescence follows a negative binomial distribution~\cite{trost2023speckle} (for details see \aref{sec:contrast}). The distribution is parametrized by the so-called speckle contrast $\beta$, which can be interpreted as a measure of the effective number of mutually incoherent modes \textemdash $\beta$ equals \num{1} for fully coherent light, and approaches \num{0} for incoherent light. In line with the standard methodology in the literature~\cite{inoue2019determination,trost2023imaging}, we can then estimate the effective pulse duration as
\begin{equation}
    T\approx \frac{5}{27} \frac{\tau_\text{c}}{\beta}.
    \label{eq:pulse_duration}
\end{equation}
The speckle contrast can be estimated for single exposures according to \(\beta = \left(\Var(I(\mathbf{k})) - \mu\right)/\mu^{2}\), where $\mu$ is the average photon count, and $\Var(I(\mathbf{k}))$ is the variance of the photon counts on the detector~\cite{trost2023speckle}.

\begin{figure}
  \centering
  \includegraphics[width=\linewidth]{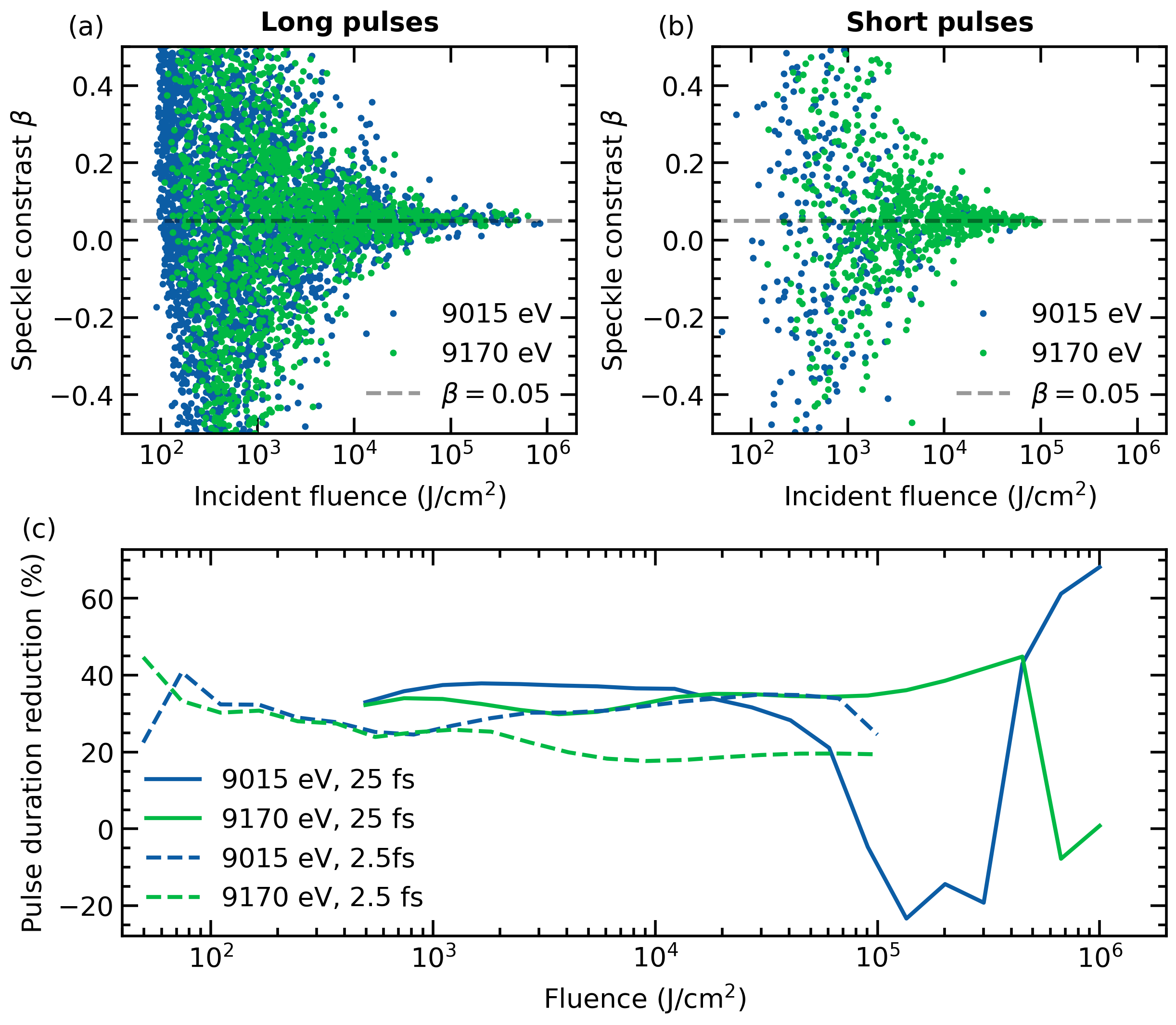}
  \caption{\label{fig:beta_nu}Speckle contrast of the K$\alpha$ fluorescence as function of fluorescent signal per pixel $\mu$ for (a)~long and (b)~short pulses at \SI{9015}{\electronvolt} and \SI{9170}{\electronvolt}. (d)~Theoretical pulse duration reduction obtained by Gaussian fits (comparing the full width at half maximum) to the fluorescence profiles over the incident pulse duration for long (solid lines) and short (dashed lines) pulse durations.}
\end{figure}

\fref{fig:beta_nu}(a,b) illustrate the speckle contrast as a function of $\mu$ at \SI{9015}{\electronvolt} and \SI{9170}{\electronvolt} for both short and long pulses. For low $\mu$, the rarity of multiple photons per pixel introduces a significant uncertainty in the estimation of the speckle contrast. As the signal level increases, the speckle contrast converges to $\beta=0.0465$ and \(0.053\) for short pulses at \SI{9015}{\electronvolt} and \SI{9170}{\electronvolt}, respectively. The speckle contrast is almost the same for the long pulses, namely $\beta=0.052$ and \(0.053\) at the same photon energies. {Although this seems surprising at first, this behaviour can be understood when considering the spiky nature of the SASE pulses. Despite being ten times longer, the long pulses had ten times higher pulse energy compared to short pulses, on average resulting in the same energy per spike in both cases.}

Together with copper K\(\alpha\) coherence time, we obtain an effective pulse duration of \SI{1.8}{\femto\second}, which is in good agreement with the pulse durations retrieved from calculations of the \(g^{(2)}(\Delta E)\). This pulse duration due to self-gating corresponds to a reduction by $\SI{28}{\percent}$ for the short pulses (from the simulated {\SI{2.5}{\femto\second}} to \SI{1.8}{\femto\second}). For long pulses, the reduction is by $\SI{93}{\percent}$, which is more significant than the so-far reported $\SI{35}{\percent}$ by \textcite{inoue2021shortening} for bulk foils using a \SI{7}{\femto\second} duration incident pulse. These conclusions are qualitatively supported by our simulations. \fref{fig:beta_nu}(c) shows the reduction in pulse duration based on Gaussian fits to the fluorescence time profiles relative to the incident pulse duration. Here, a negative number means an elongation in the pulse duration. For long pulses (solid lines), theory predicts the reduction is around \SIrange{30}{40}{\percent} for fluence values \SI{<1E5}{\joule\per\centi\meter\squared}. At the highest fluences simulated and for \SI{9015}{\electronvolt} photon energy, there is first a phase where pulses get longer (shown as negative reduction), followed by a reduction of about \SI{50}{\percent}. For short pulses (dashed lines), the reduction stays constant at a \SIrange{20}{40}{\percent}.

In summary, we observed saturable absorption and reverse saturable absorption in single copper nanoparticles. Our results confirm that, on femtosecond timescales, the nonlinear response primarily depends only on the incident fluence and not on the pulse duration. Furthermore, from our analysis of the coherence properties of the fluorescent emission, we report the same speckle contrast of $\beta\approx0.05$ for the short (nominal \SI{2.5}{\femto\second}) and long (nominal \SI{25}{\femto\second}) XFEL pulses. In line with the reasoning of \citet{inoue2019determination} and \textcite{trost2023speckle}, this would correspond to pulse shortening due to self-gating by \SI{28}{\percent} and by up to \SI{93}{\percent} for the short and long pulses, respectively. {These results also indicate that the important parameters that lead to the reported changes in the absorption cross section are the fluence per FEL spike and the number of spikes, as both pulse durations investigated here provided the same fluence per spike.} Below the edge, we observed the onset of K\(\alpha\) fluorescence owing to the K-M resonant excitation when copper reaches high ionization~\cite{cardoch2023decreasing}. By detecting both coherent diffraction and fluorescence emission, our single particle experimental setup opens the possibility of studying resonance-induced nonlinear diffraction enhancement from different materials. Our results provide a quantitative benchmark for simulations of single nanoparticles in the high-fluence regime, and they establish the temporal window and fluence conditions under which fluorescence-based imaging operates; in the companion Letter~\cite{wollweber2025} we exploit these conditions to realise single-particle IDI and identify ASE-induced limits to its visibility. Transient resonances have been shown to enhance diffraction imaging of xenon clusters in the soft X-ray regime~\cite{kuschel_nonlinear_2025}. The short pulse durations employed here are necessary to beat the onset of ionic damage. Future experiments could focus on high-resolution coherent diffraction to observe changes in atomic structure factors.
\begin{acknowledgments}
  \emph{Acknowledgments}\textemdash
  We acknowledge European XFEL in Schenefeld, Germany, for provision of X-ray free-electron laser beamtime at SPB/SFX SASE1 under proposal number 5476 and would like to thank the staff for their assistance. 
  This work is supported by the Cluster of Excellence `CUI: Advanced Imaging of Matter' of the Deutsche Forschungsgemeinschaft (DFG) - EXC 2056 - project ID 390715994. %
  S.\ C.\ and N.\ T.\ acknowledge the Swedish Research Council (Grants No. 2019-03935 and No. 2023-03900) for financial support. %
  C.\ C.\ acknowledges the Swedish Research Council (Grant No. 2018-00740), the Röntgen-Ångström Cluster (Grant No. 2019-03935), and the Helmholtz Association through the Center for Free-Electron Laser Science at DESY. %
  S.K and J.v.Z acknowledge funding by the Deutsche Forschungsgemeinschaft within the TRR 306 QuCoLiMa (“Quantum Cooperativity of Light and Matter”), project-ID 429529648.
  This research was supported in part through the Maxwell computational resources operated at Deutsches Elektronen-Synchrotron DESY, Hamburg, Germany. %
  Calculations were performed on the Davinci computer cluster provided by the Laboratory of Molecular Biophysics at Uppsala University. %
\end{acknowledgments}
\bibliography{lib}
\appendix
\section{Estimate of total fluorescence emission}
\label{sec:sigma}
To calculate the fluorescence shown in \fref{fig:exp_above_edge}, we extrapolate the measured fluorescent intensity in the shadow of the nickel filter to $4\pi \, \text{sr}$, assuming uniform isotropic emission. Furthermore, we correct for the angular dependent absorption $A$ of the nickel filter according to
\begin{equation}
  A = 1 - \exp\left(\frac{-\ell}{L\,cos(\theta)}\right),
\end{equation}
where $\ell=$\SI{20}{\micro\meter} is the thickness of the nickel filter and $L=$\SI{23.95}{\micro\meter} is its absorption length at \SI{90}{\degree} and photon energy of $\SI{8040}{\electronvolt}$. The fluorescence cross section $\sigma_{\text{atom}}$ is then given by
\begin{equation}
  \sigma_{\text{atom}} = \frac{E_{4\pi}}{N\,\Phi},
\end{equation}
where $E_{4\pi}$ is the total fluorescence energy emitted in $4\pi$, $N$ is the number of atoms in the particle and $\Phi$ is the incident fluence.
\section{\label{sec:modelling}Plasma modelling}
The collisional radiative code solves a rate matrix to determine changes in the copper's electronic occupation over time. The Atomic data used to construct the rate matrix extends to $n\leq3,\ell\leq2$ and \(4s\) electronic states with at most one excitation from the K and two from the L and M shells, and includes photoionization and radiative recombination, photoexcitation and radiative de-excitation, autoionization and dielectronic recombination, as well as electron impact ionization and three-body recombination. Collisional excitation and de-excitation cross sections are computed from oscillator strengths using the Bethe approximation~\cite{scott_advances_2010}. The code fits collisional cross sections as a function of impact energy. Free electron and ion temperatures follow a Maxwellian distribution. We incorporate effects of electron degeneracy using a factor that scales collisional impact ionization and excitation cross sections (\citet[Eq.~4.48 and Eq.~4.51]{scott_collisional-radiative_2016}, respectively)~\cite{tallents_free_2016}. We use a Steward-Pyatt continuum lowering model~\cite{stewart_lowering_1966} that is scaled by a factor of 0.53 to achieve +1 ionization, approximating copper's conduction band at room temperature, at the beginning of the simulation. Photon energies are shifted by \SI{17}{\electronvolt} to match the atomic model's K edge to its tabulated value~\cite{henke1993x}. Radiation transport is solved over a coarse energy grid consisting of 100 logarithmically-spaced points that span from \SIrange{0.1}{1E4}{\electronvolt}. To resolve finer details around copper's K-edge, we include an additional 100 logarithmically-spaced grid between \SIrange{7.8}{9.2}{\kilo\electronvolt}. From converged solutions, we compute high-resolution spectra around the K$\alpha$ emission region.
\section{\label{sec:contrast}Degree of coherence}
The fluorescence has more of the characteristics of coherent laser-like light, the shorter the pulse, since more of the emitted photons can interfere with each other. One way to estimate the degree of coherence is by using the distribution of observed photon counts per pixel. The photon statistics for partially coherent light follow a negative binomial distribution~\cite{goodman2007speckle,trost2023speckle}, according to which, the probability of measuring $x$ photons in a pixel is given by
\begin{equation}
  P_{\text{NB}}(x|\mu,\beta) = \frac{(\beta\mu)^x(1+\beta\mu)^{-(1+\beta x)/\beta} \, \Gamma(x+1/\beta)}{x! \, \Gamma(1/\beta)}.
\end{equation}
\end{document}